\documentclass[12pt]{article}
\usepackage{amsmath,amssymb}
\usepackage{bm}
\usepackage{graphicx,color}
\usepackage{wrapfig}
\begin{document}
\title{
\begin{flushright}
\ \\*[-80pt] 
\begin{minipage}{0.2\linewidth}
\normalsize
\end{minipage}
\end{flushright}
{\Large \bf Linking Leptonic  CP violation \\
 to  Quark Unitarity Triangle
\\*[20pt]}}

\author{
\centerline{
Morimitsu~Tanimoto\footnote{E-mail address: tanimoto@muse.sc.niigata-u.ac.jp} \ \ 
and \ \ Kei Yamamoto\footnote{E-mail address: yamamoto@muse.sc.niigata-u.ac.jp}}
\\*[20pt]
\centerline{
\begin{minipage}{\linewidth}
\begin{center}
{\it \normalsize
Department of Physics, Niigata University,~Niigata 950-2181, Japan }
\end{center}
\end{minipage}}
\\*[70pt]}

\date{
\centerline{\small \bf Abstract}
\begin{minipage}{0.9\linewidth}
\vskip  1 cm
\small
We study the linking between the CP violating phase of the  lepton sectors
and  the unitarity triangle of the $B^0$ meson system.
Antusch, King, Malinsky and Spinrath have shown
that the quark mass matrices with the negligible 1-3 mixing 
give an interesting relation
 between the  phase of the quark mixing matrices and  CP violating measure $\phi_2(\alpha)$. 
This approach is extended  by considering 
 the $SO(10)$ GUT including the Pati-Salam symmetry, which
links the Yukawa matrices of the quark sector  to the one of the lepton sector. 
We discuss the relation of the CP violating phases between both quark and lepton sectors
as well as the mixing angles.
Then, the leptonic CP violating phase is predicted
in terms of the angle of the unitarity triangle of the $B^0$ meson system.
 The leptonic CP violating phase $\delta_{\rm PMNS}$ is predicted
in the region $-74^\circ\sim -89^\circ$,
 which is the consistent with the recent  T2K results.
Our predicted phase is sensitive to  $\phi_2(\alpha)$ and $\phi_3(\gamma)$.
These  predictions will be clearly tested in the future neutrino experiments
as well as the Belle-II experiment.
\end{minipage}
}

\begin{titlepage}
\maketitle
\thispagestyle{empty}
\end{titlepage}

\section{Introduction}

\vskip 1 cm

The neutrino oscillation experiments are going on the new step to reveal the CP violation
in the lepton sector.
The T2K experiment has confirmed the neutrino oscillation in the $\nu_\mu \to\nu_e$ appearance events 
\cite{Abe:2013hdq}, 
which provide us the new information of the CP violation of the lepton sector by combining  the data of reactor experiments ~\cite{An:2012eh,Ahn:2012nd,Adamson:2013whj,Abe:2014lus}.
Therefore, the detailed study of the neutrino mixing including the CP violating phase
gives us clues to reach  the  flavor theory 
\cite{Kang:2000sb}-\cite{Kang:2015xfa}.

On the other hand, the CP violating phase is well determined in the quark sector since
the unitarity triangle of the $B^0$ meson system has been confirmed.
The measurements of  the three angles $\phi_1(\beta)$,  $\phi_2(\alpha)$
and  $\phi_3(\gamma)$ will be  improved considerably by the Belle-II experiment.
In the extensions of the standard model (SM), 
the grand unified theories (GUTs) link the Yukawa matrices of the quark sector 
to that of the lepton sector. 
Therefore, one can study  the  relation of the CP violating phases between the  quark and the lepton sectors as well as the mixing angles.

The Yukawa matrices which have the  1-3 texture zeros (negligible 1-3 mixing) 
in both up- and down-quark sectors 
give us a clear relation 
between the Yukawa phase and  the CP violating measure $\phi_2(\alpha)$
as discussed by  Antusch, King, Malinsky and Spinrath \cite{Antusch:2009hq}.
In this paper, we extend this approach of the CP violating phase  to the lepton sector
\cite{Antusch:2012fb,Antusch:2013kna} and discuss the link of the CP violating phase
between the quark and lepton sectors. 

By using some GUTs, the reactor angle of  neutrinos  is  related to the
quark mixing angle \cite{Antusch:2011qg,Marzocca:2011dh,BhupalDev:2011gi,BhupalDev:2012nm}.
 Actually, one of the authors has examined  the mixing angles
of both quarks  and the leptons  in the  $SO(10)$ GUT including the Pati-Salam symmetry
\cite{Bando:2003wb,Bando:2004hi}.
In this paper,  we  predict the leptonic CP violating phase 
in terms of the unitarity triangle of  quarks by using
the $SO(10)$ GUT with the Pati-Salam symmetry.
The leptonic CP violating phase  $\delta_{\rm PMNS}$ is predicted to be $-74^\circ\sim -89^\circ$,
which is the opposite sign against the quark CP violating phase  $\delta_{\rm CKM}$.
 This prediction  is  consistent with the recent  T2K results \cite{Abe:2013hdq}.

The paper is organized as follows. In section 2,
we present  the unitarity triangle, $\phi_1(\beta)$,  $\phi_2(\alpha)$, $\phi_3(\gamma)$ and $J_{CP}$
in terms of Yukawa phases, and examine these relations numerically by inputing the recent 
experimental data of the CKM matrix.
The  CP violating phase of the lepton sector is also discussed.
In section~3, we discuss the linking between the quark CP violating phase and
the leptonic  CP violating phase by considering the $SO(10)$ GUT, and predict the magnitude of the leptonic  CP violating phase in terms of the angle of the unitarity triangle. 
The Section 4 is devoted to the summary. 
In the Appendix, we present the general formula of the quark and the lepton mixing.


\section{Mixing sum rules in quarks and leptons}

\subsection{Quark mixing angle sum rules}

 Let us start  with discussing the quark mixing angle sum rules
presented by Antusch, King, Malinsky and Spinrath \cite{Antusch:2009hq}.
  The quark mass matrices are given in the Yukawa sector as:
 
 \begin{eqnarray}
 {\cal L}_Y= -\bar{u^i_L} (M_u)_{ij} u_R^j -  \bar{d^i_L} (M_d)_{ij} d_R^j \ ,
 \end{eqnarray}
 where $M_u$ and $M_d$ are the mass matrices of the up- and the down-quarks, respectively.
  The mass matrices are diagonalized
 
 \begin{eqnarray}
  V_{uL}^\dagger M_u V_{uR} = {\rm diag} (m_u, m_c, m_t) \ , \qquad
  V_{dL}^\dagger M_d V_{dR} = {\rm diag} (m_d, m_s, m_b) \ , 
 \end{eqnarray}
 where $ V_{uL}$, $V_{uR}$,  $V_{dL}$ and   $V_{dR}$ are unitary $3\times 3$ matrices.
 The CKM matrix $U'_{\rm CKM}$ is given
 \begin{eqnarray}
 U'_{\rm CKM} = V_{uL}^\dagger V_{dL},
 \end{eqnarray}
where the unphysical phases are included.


The unitary matrix $V_{qL}\ (q=u,d)$ can be written in terms of three angles 
$\theta_{ij}^q$ and three phases $\delta_{ij}^q$ as:
\begin{eqnarray}
V_{qL} = U_{23}^{qL} U_{13}^{qL} U_{12}^{qL} 
 ,
\end{eqnarray}
where $U_{12}^{qL}$, $U_{23}^{qL}$ and $U_{13}^{qL}$ are given as,
\begin{align}
U_{12}^{qL}=
\begin{pmatrix}
c_{12}^q  & s_{12}^q e^{-i\delta_{12}^q} &0 \\
-s_{12}^q e^{i\delta_{12}^q}&  c_{12}^q &0 \\    0 &  0 & 1
\end{pmatrix},  \quad
U_{23}^{qL}=
\begin{pmatrix}
1 &0 &0 \\
0 & c_{23}^q & s_{23}^q e^{-i\delta_{23}^q} \\
0 & -s_{23}^q e^{i\delta_{23}^q}&  c_{23}^q  
\end{pmatrix},
\label{UCKM}
\end{align}
 and  $U_{13}^{qL}$ is also analogously written.
Here $c_{ij}^q$ $s_{ij}^q$ are abbreviations for $\cos\theta_{ij}^q$
and  $\sin\theta_{ij}^q$, where $\theta_{ij}^q$ is always made positive by
the suitable choice of the phase $\delta_{ij}^q$.
Then, we can express the CKM matrix in terms of the up- and the down-mixing matrices as:
\begin{eqnarray}
U'_{\rm CKM} =U_{12}^{uL\dagger} U_{13}^{uL\dagger}U_{23}^{uL\dagger}
   U_{23}^{dL} U_{13}^{dL} U_{12}^{dL} \ .
\end{eqnarray}

We can also express the CKM matrix  by the products of three unitary matrices
 as follows:
\begin{eqnarray}
U'_{\rm CKM}= U_{23} U_{13} U_{12} \ ,
\label{CKMprime}
\end{eqnarray}
where
\begin{align}
U_{12}=
\begin{pmatrix}
c_{12}  & s_{12} e^{-i\delta_{12}} &0 \\
-s_{12} e^{i\delta_{12}}&  c_{12} &0 \\    0 &  0 & 1
\end{pmatrix},
\label{CKM12}
\end{align}
and so on. Three mixing angles $\theta_{ij}$ and  three phases $\delta_{ij}$ appear
in $U'_{\rm CKM}$ of Eq.(\ref{CKMprime}),
but two of three phases are removed away by multiplying  the phase matrix
in the right-hand side.
On the other hand, in the PDG parametrization~\cite{Beringer:1900zz},
the CKM matrix is given in terms of   three mixing angles  $\theta_{ij}$,
which are same ones in Eq.(\ref{CKM12}),
and  one  phase   $\delta _{\rm CKM}$ as follows:
\begin{align}
U_{\rm CKM}=
\begin{pmatrix}
c_{12} c_{13} & s_{12} c_{13} & s_{13}e^{-i\delta _{\rm CKM}} \\
-s_{12} c_{23} - c_{12} s_{23} s_{13}e^{i\delta _{\rm CKM}} & 
c_{12} c_{23} - s_{12} s_{23} s_{13}e^{i\delta _{\rm CKM}} & s_{23} c_{13} \\
s_{12} s_{23} - c_{12} c_{23} s_{13}e^{i\delta _{\rm CKM}} & 
-c_{12} s_{23} - s_{12} c_{23} s_{13}e^{i\delta _{\rm CKM}} & c_{23} c_{13} \ 
\end{pmatrix},
\label{mixing}
\end{align}
where  $c_{ij}$ and   $s_{ij}$  denote  $\cos\theta_{ij}$
and  $\sin\theta_{ij}$, respectively. 
 Consequently, the  CP violating phase $\delta_{\rm CKM}$ is expressed
in terms of $\delta_{12}$, $\delta_{23}$ and $\delta_{13}$  as:
\begin{eqnarray}
\delta_{\rm CKM}=\delta_{13}-\delta_{23}-\delta_{12} \ .
\label{deltaCKM}
\end{eqnarray}


Let us consider the phenomenological viable textures for the up- and the down-quark mass matrices,
 which have
 the  1-3 texture zeros (negligible 1-3  mixing).
Therefore, taking  $\theta_{13}^d=\theta_{13}^u=0$, we can express
the CKM matrix
\begin{eqnarray}
U'_{\rm CKM} =U_{12}^{uL\dagger} U_{23}^{uL\dagger} U_{23}^{dL} U_{12}^{dL},
\label{U13zero}
\end{eqnarray}
where $U_{13}^{qL}$ is the unit matrix.
By using Eqs.(\ref{CKMprime}) and (\ref{U13zero}),
we can derive the following relations,
\begin{eqnarray}
&&\theta_{23} e^{-i\delta_{23}}=\theta_{23}^d e^{-i\delta_{23}^d}
  -\theta_{23}^u e^{-i\delta_{23}^u}  \ , 
\label{qmix23} \\
&&\theta_{13} e^{-i\delta_{13}}=-\theta_{12}^u  e^{-i\delta_{12}^u}
(\theta_{23}^d e^{-i\delta_{23}^d} -\theta_{23}^u e^{-i\delta_{23}^u})  \ , 
\label{qmix13} \\
&&\theta_{12} e^{-i\delta_{12}}=\theta_{12}^d e^{-i\delta_{12}^d}
  -\theta_{12}^u e^{-i\delta_{12}^u}  \ , 
\label{qmix12}
\end{eqnarray}
where we take the leading order in the small mixing angle
 by putting  $c_{ij}^{u,d}\simeq 1$ and  $s_{ij}^{u,d}\simeq \theta^{u,d}_{ij}$.
 In order to see the effect of
non-vanishing  $\theta_{13}^u$ and  $\theta_{13}^u$, we present 
the general formula and discussions in the Appendix.

By using Eqs.(\ref{qmix23}) and (\ref{qmix13}), one obtains
\begin{eqnarray}
 \theta^u_{12} = \frac{\theta_{13}}{\theta_{23}} \ .
   \label{u12}
\end{eqnarray}
By combining Eqs.(\ref{qmix23}), (\ref{qmix13}) and (\ref{qmix12}),
 one gets
 \begin{eqnarray}
 \theta_{12}^d  e^{-i(\delta_{12}^d-\delta_{12})}
=\theta_{12} - \frac{\theta_{13}}{\theta_{23}} 
  e^{-i(\delta_{13}-\delta_{23}-\delta_{12})} \ ,
\end{eqnarray}
which gives
\begin{eqnarray}
 \theta_{12}^d 
=\left |\theta_{12} - \frac{\theta_{13}}{\theta_{23}} 
  e^{-i \delta_{\rm CKM}} \right |   \ .
  \label{d12}
\end{eqnarray}

One also gets a relation by using Eqs. (\ref{qmix23}), (\ref{qmix13}) and (\ref{qmix12}):
\begin{eqnarray}
 \frac{\theta_{13}   \theta_{12} }{ \theta_{23}} e^{i \delta_{\rm CKM}}
= -\theta_{12}^u [\theta_{12}^d e^{-i(\delta_{12}^d-\delta_{12}^u)} - \theta_{12}^u]  \ .
\end{eqnarray}
Therefore, one obtains  the phase sum rule \cite{Antusch:2009hq}
\begin{eqnarray}
 \delta_{\rm CKM}={\rm Arg} \left[1- \frac{\theta_{12}^d}{\theta_{12}^u}
 e^{-i(\delta_{12}^d-\delta_{12}^u)}  \right ] \ ,
 \label{delta12}
\end{eqnarray}
which is just the angle $\phi_3 (\gamma)$ of the unitarity triangle as seen  later.

The three angles of the unitarity triangle 
$\phi_1(\beta)$, $\phi_2(\alpha)$, $\phi_3(\gamma)$  can be expressed 
in terms of  $\delta_{12}^d-\delta_{12}^u$ as follows:

\begin{eqnarray}
&&\phi_1 (\beta) = {\text Arg} \left [-\frac{U_{cd}U_{cb}^*}{U_{td}U_{tb}^*} \right ]
={\text Arg} \left [  1-\frac{\theta_{12}^u}{\theta_{12}^d} 
     e^{-i(\delta_{12}^d-\delta_{12}^u)} \right ] , \nonumber \\
&&\phi_2 (\alpha) = {\text Arg} \left [-\frac{U_{td}U_{tb}^*}{U_{ud}U_{ub}^*} \right ]
=\delta_{12}^d-\delta_{12}^u , \nonumber \\   
&&\phi_3 (\gamma) = {\text Arg} \left [-\frac{U_{ud}U_{ub}^*}{U_{cd}U_{cb}^*} \right ]
={\text Arg} \left [  1-\frac{\theta_{12}^d}{\theta_{12}^u} 
     e^{-i(\delta_{12}^d-\delta_{12}^u)} \right ] ,
\label{triangle}
\end{eqnarray}
where the CKM matrix elements  $U_{ij}$'s  are expressed
in terms of $\theta^q_{ij}$ and $\delta^q_{ij}$ by using Eq.(\ref{U13zero}).
We can easily check $\phi_1+\phi_2+\phi_3=\pi$.
It is noticed that $\phi_3 (\gamma)$ is just $\delta_{\rm CKM}$.
Thus, the CP violating phases are given
by  $\delta_{12}^d-\delta_{12}^u$ and $\theta_{12}^d/\theta_{12}^u$ in this scheme.
We can also give another CP violating measure, the Jarlskog invariant $J_{CP}$~\cite{Jarlskog:1985ht},  as;
\begin{equation}
J_{CP}=\text{Im}\left [U_{us}U_{cb}U_{ub}^\ast U_{cs}^\ast \right ]
=|U_{cb}|^2  \theta_{12}^u \theta_{12}^d \sin (\delta_{12}^d - \delta_{12}^u) \ .
\label{JCP}
\end{equation}
\begin{figure}[b!]
\begin{minipage}[]{0.45\linewidth}
\vspace{4 mm}
\includegraphics[width=7.5cm]{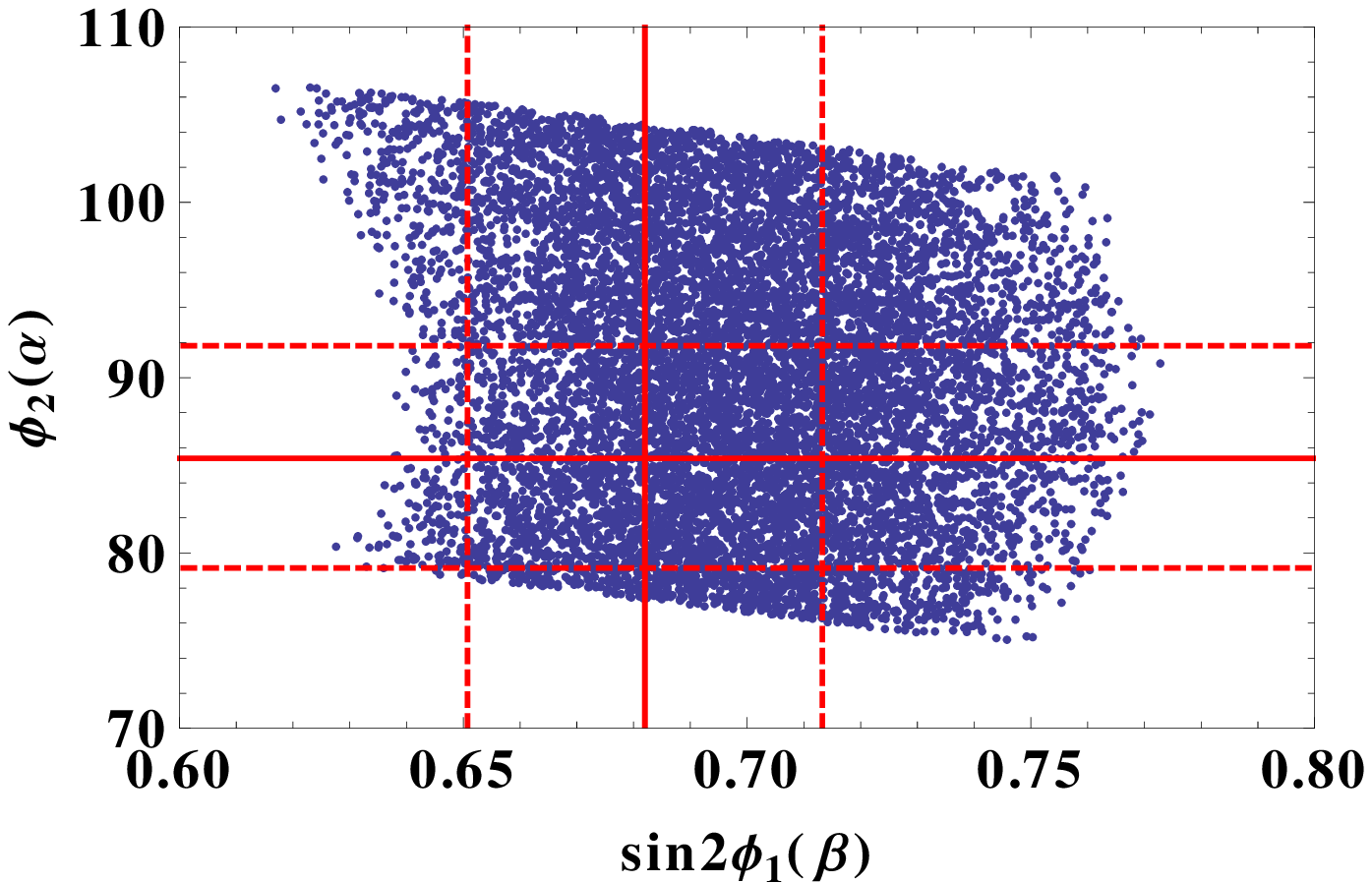}
\caption{The predicted region on the $\sin 2\phi_1(\beta)$-$\phi_2(\alpha)$ plane.
The solid   and dashed lines denote the central values and
the errors with $90\%$ C.L.  of the experimental data, respectively.}
\label{fig1}
\end{minipage}
\hspace{5mm}
\begin{minipage}[]{0.45\linewidth}
\vskip - 3 mm
\includegraphics[width=7.5cm]{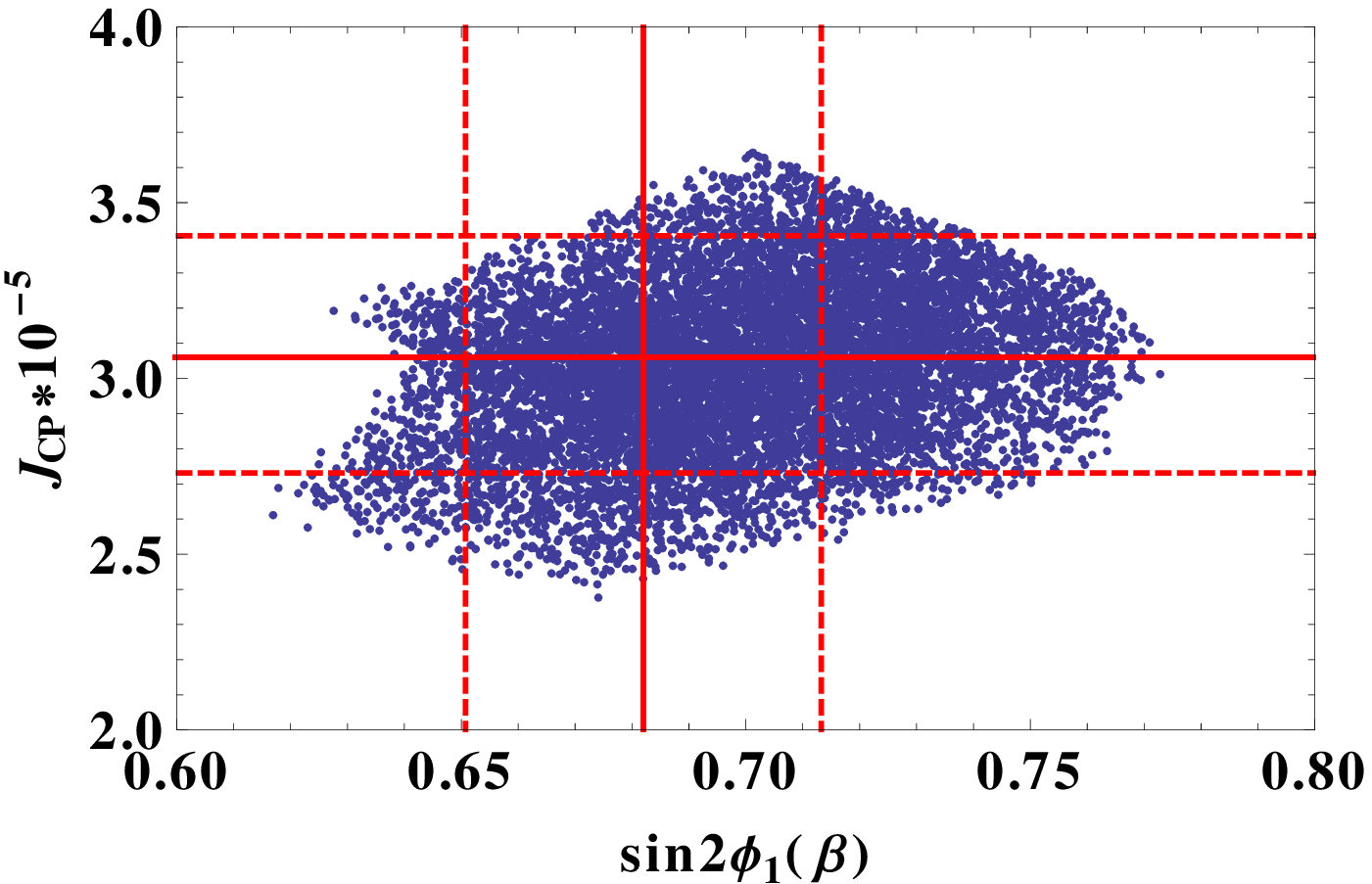}
\caption{The predicted region on the $\sin 2\phi_1(\beta)$-$J_{CP}$ plane.
The solid   and dashed lines denote the central values and
the errors with $90\%$ C.L.  of the experimental data, respectively.}
\label{fig2}
\end{minipage}
\end{figure}
\begin{figure}[t!]
\begin{minipage}[]{0.45\linewidth}
\vspace{4 mm}
\includegraphics[width=7.5cm]{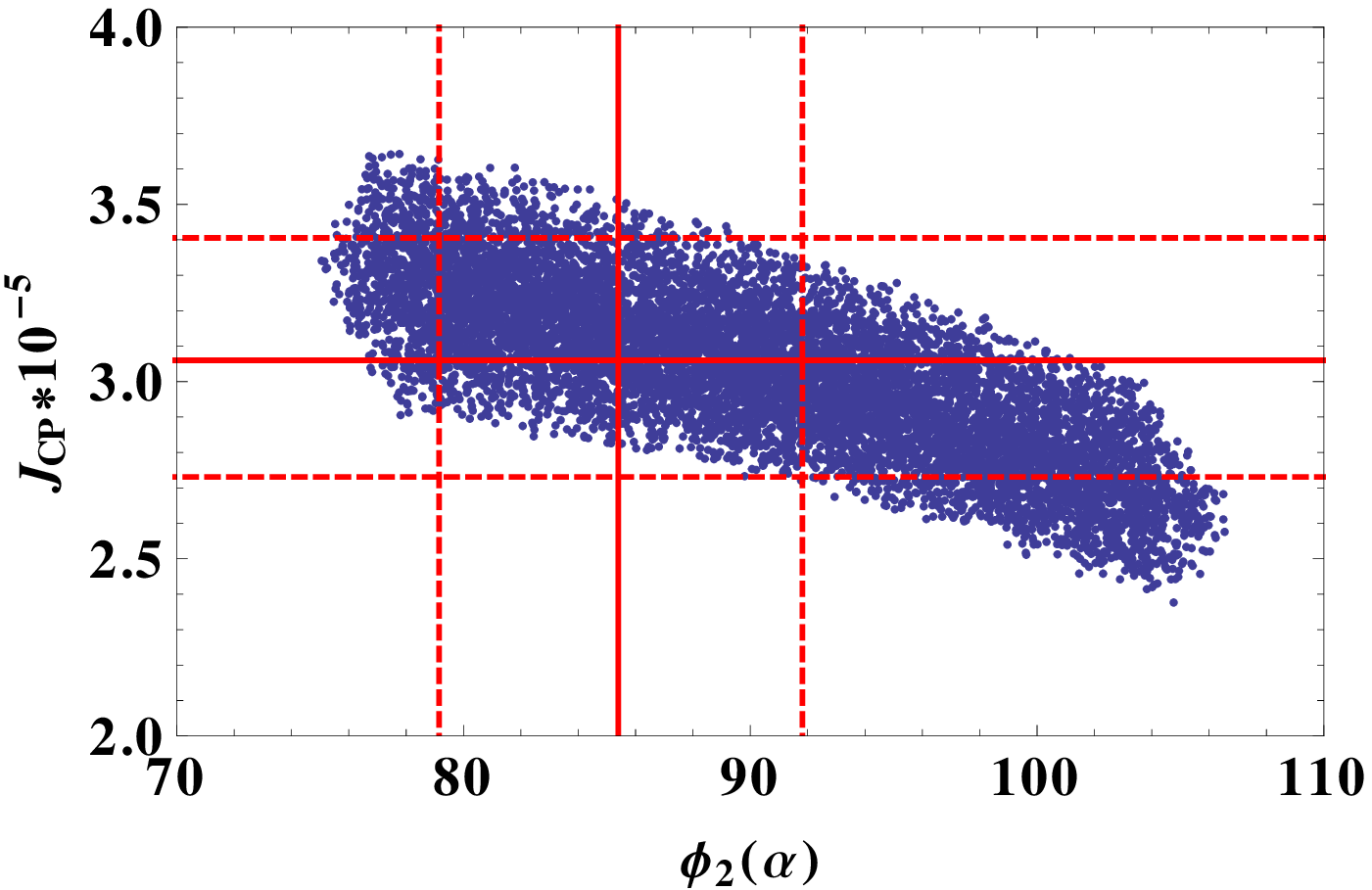}
\caption{The predicted region on the $\phi_2(\alpha)$-$J_{CP}$ plane.
The solid   and dashed lines are same in Figs. 1 and 2.}
\label{fig3}
\end{minipage}
\hspace{5mm}
\begin{minipage}[]{0.45\linewidth}
\vskip - 3 mm
\includegraphics[width=7.5cm]{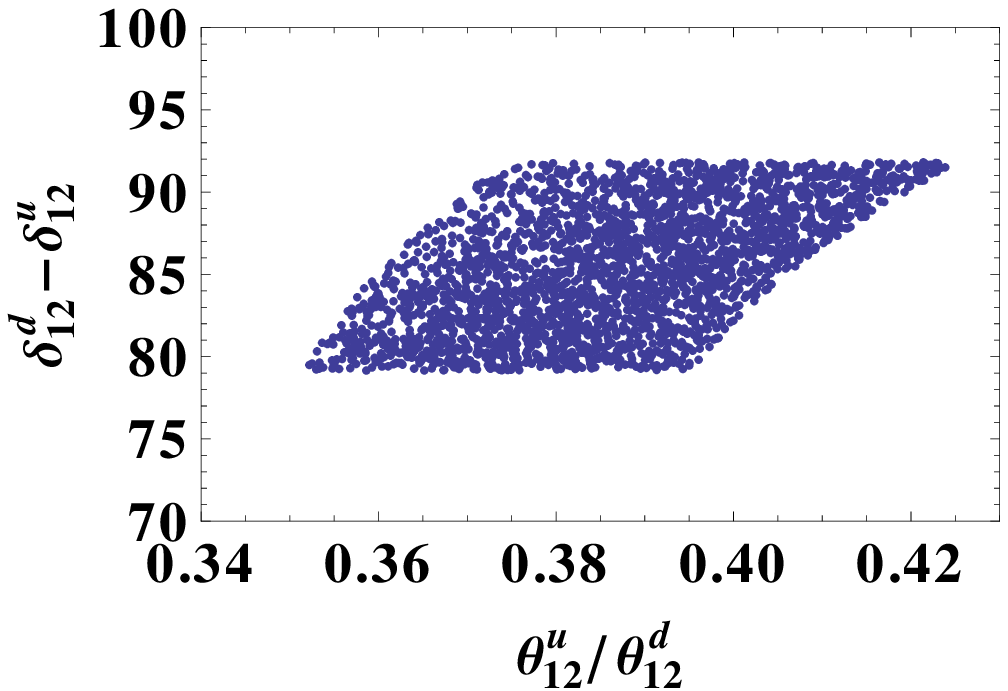}
\caption{The allowed region of  $\delta_{12}^d-\delta_{12}^u$ and $\theta_{12}^d/\theta_{12}^u$  with the constraint of the unitarity triangle.}
\label{fig4}
\end{minipage}
\end{figure}

Let us show numerical results in order to test the consistency of this scheme.
We input the   PDG data ~\cite{Beringer:1900zz} 
for the CKM mixing elements and phase as follows:
\begin{eqnarray}
&&
 |U_{us}|=0.22536\pm 0.00061 \ , \quad 
 |U_{ub}|=0.00355\pm  0.00015 \ , \quad
 |U_{cb}|=0.0414\pm 0.0012 \ ,  \nonumber \\ 
&& \delta_{CKM}=\phi_3(\gamma)={68.0^\circ} ^{+8.0^\circ }_{-8.5^\circ }\ .
\end{eqnarray}
Then we can obtain  $\delta_{12}^d-\delta_{12}^u$ and $\theta_{12}^d/\theta_{12}^u$
by using Eqs. (\ref{u12}), (\ref{d12}) and (\ref{delta12}),
and so we can calculate   $\sin 2\phi_1(\beta)$,  $\phi_2(\alpha)$ and $J_{CP}$
by using Eqs. (\ref{triangle}) and (\ref{JCP}).

Taking  account of the experimental error with $90\%$ C.L. for the  input data,
we plot the calculated regions of $\sin 2\phi_1(\beta)$,  $\phi_2(\alpha)$ and $J_{CP}$
 on  the planes of 
 $\sin 2\phi_1(\beta)$-$\phi_2(\alpha)$,  $\sin 2\phi_1(\beta)$-$J_{CP}$ and 
 $\phi_2(\alpha)$-$J_{CP}$ in Figures 1, 2 and 3, respectively.
In these figures, 
the following  experimental bounds   with $90\%$ C.L.  are also shown:
\begin{eqnarray}
\sin 2\phi_1(\beta)=0.682\pm 0.019, \quad
\phi_2(\alpha)={85.4^\circ} ^{+3.9^\circ }_{-3.8^\circ }\ , \quad 
J_{CP}={3.06} ^{+0.21}_{-0.20} \times 10^{-5}\ . 
\label{data}
\end{eqnarray}
The calculated  $\sin 2\phi_1(\beta)$, $\phi_2(\alpha)$ and  $J_{CP}$ are consistent with the experimental data
although some  regions are excluded by the experimental bounds.
 Thus, the CP violation is successfully  expressed in terms of
$\delta_{12}^d-\delta_{12}^u$ and $\theta_{12}^d/\theta_{12}^u$,
which come from the quark mass matrices.
By imposing constraints of the experimental data of the unitarity triangle in  Eq.(\ref{data}),
we finally obtain the allowed region of  $\delta_{12}^d-\delta_{12}^u$ and $\theta_{12}^d/\theta_{12}^u$ as shown in Figure 4.
Taking this allowed region of $\delta_{12}^d-\delta_{12}^u$,
 we will discuss the CP violating phase in the lepton sector.


\subsection{Lepton mixing angle sum rules}

Similar  discussions lead to  sum rules for the mixing angles and the CP violating phase  
in the lepton sector \cite{Antusch:2012fb,Antusch:2013kna}.
For the lepton mixing \cite{Maki:1962mu,Pontecorvo:1967fh},
we write a general unitarity matrix $U'_{\rm PMNS}$
including  the Majorana phases $\varphi_1$, $\varphi_2$ and
the Dirac phase $\delta_{\rm PMNS}$ as follows:
\begin{eqnarray}
U'_{\rm PMNS} ={\rm diag}(e^{i\delta_e},e^{i\delta_\mu},e^{i\delta_\tau})\cdot
U_{\rm PMNS}\cdot {\rm diag}(e^{-i\varphi_1/2},e^{-i\varphi_2/2},1) \ .
\end{eqnarray}
Here $U_{\rm PMNS}$ is the lepton mixing matrix corresponding to the quark one of Eq.(\ref{mixing}),
in which $\theta_{ij}$ and  $\delta_{ij}$ are redefined in the lepton sector.
We replace the quark mixing angles
and the  phases $\theta_{ij}^d$, $\theta_{ij}^u$, $\delta_{ij}^d$, $\delta_{ij}^u$
with  the lepton ones $\theta_{ij}^e$, $\theta_{ij}^\nu$,  $\delta_{ij}^e$, $\delta_{ij}^\nu$, 
where superscripts $e$ and $\nu$ correspond to the charged lepton and 
the neutrino, respectively.

The observed  mixing angles $\theta_{23}$ and $\theta_{12}$ are order one
 while $\theta_{13}$  is order of the Cabibbo angle in the experimental data.
Therefore, it is reasonable to suppose  $\theta_{13}^e=\theta_{13}^\nu=0$
as well as in the quark sector. 
So,  we  can take the PNMS matrix    to be
\begin{eqnarray}
U'_{\rm PMNS} =U_{12}^{e L\dagger} U_{23}^{e L\dagger} U_{23}^{\nu L} U_{12}^{\nu L} \ .
\end{eqnarray}
Then, the PMNS mixing angles $\theta_{ij}$'s are given in terms of 
  $\theta_{ij}^e$, $\theta_{ij}^\nu$, 
and  $\delta_{ij}^e$, $\delta_{ij}^\nu$ as follows:
\begin{eqnarray}
&&c_{13}s_{23} e^{-i\delta_{23}}=s_{23}^\nu e^{-i\delta_{23}^\nu} 
-\theta_{23}^e c_{23}^\nu e^{-i\delta_{23}^e}  \ , 
\label{lmix23} \\
&&\theta_{13} e^{-i\delta_{13}}=-\theta_{12}^e  e^{-i\delta_{12}^e}
(s_{23}^\nu e^{-i\delta_{23}^\nu} -\theta_{23}^e c_{23}^\nu e^{-i\delta_{23}^e})  \ , 
\label{lmix13} \\
&&c_{13}s_{12} e^{-i\delta_{12}}=s_{12}^\nu e^{-i\delta_{12}^\nu}
  -\theta_{12}^e c_{23}^\nu c_{12}^\nu  e^{-i\delta_{12}^e}  \ , 
\label{lmix12}
\end{eqnarray}
where we take $\cos \theta_{ij}^e\simeq 1$, $\sin \theta_{ij}^e\simeq \theta_{ij}^e$
and $\sin \theta_{13}\simeq \theta_{13}$. 
The leptonic CP phases can be extracted via
\begin{eqnarray}
\delta_{12}=\frac{1}{2}(\varphi_2-\varphi_1) \ , \qquad
\delta_{23}=-\frac{1}{2}\varphi_2 \ , \qquad
\delta_{13}=\delta_{\rm PMNS}-\frac{1}{2}\varphi_1 \ .
\end{eqnarray}

In this situation,    
 the neutrino large mixing angles come from the neutrino mass matrix, and
the reactor angle $\theta_{13}$ is given by  the correction from the charged lepton sector.
By Eqs.(\ref{lmix23}) and (\ref{lmix13}), we obtain
\begin{eqnarray}
 \theta_{13} e^{-i \delta_{13}} = 
 -\theta_{12}^e c_{13} s_{23} e^{-i(\delta_{12}^e+\delta_{23})} \ ,
   \label{l13}
\end{eqnarray}
which gives us  simple sum rules for mixing angles and  phases 
\begin{eqnarray}
 \theta_{13} = \theta_{12}^e s_{23}  \ , \qquad \delta_{13} =  \delta_{12}^e+ \delta_{23}-\pi\ .
   \label{prediction13}
\end{eqnarray}
Supposing  $\theta_{23}^e\ll 0.1$ in  Eq.(\ref{lmix23}), we obtain 
\begin{eqnarray}
 s_{23} \simeq s_{23}^\nu \ , \quad  \qquad   \delta_{23} =  \delta_{23}^\nu \ .
   \label{prediction23}
\end{eqnarray}
On the other hand,
 Eq.(\ref{lmix12}) turns to
 \begin{eqnarray}
 c_{13}s_{12} e^{-i\delta_{12}}
=-\theta_{12}^e c_{23}^\nu c_{12}^\nu e^{-i\delta_{12}^e}
\left (1- \frac{s_{12}^\nu}{\theta_{12}^e c_{23}^\nu c_{12}^\nu} 
 e^{-i(\delta_{12}^\nu-\delta_{12}^e)} \right )\ , 
\end{eqnarray}
 which leads to
 \begin{eqnarray}
 s_{12}= \left |s_{12}^\nu e^{-i(\delta_{12}^\nu-\delta_{12}^e)}-\theta_{12}^e c_{23}^\nu c_{12}^\nu
  \right | \ ,
 \qquad \delta_{12} =  \delta_{12}^e-\phi-\pi ,
   \label{prediction12}
\end{eqnarray}
where $\phi$ is given by
 \begin{eqnarray}
 \phi={\rm Arg} \left[1- \frac{s_{12}^\nu}{\theta_{12}^e c_{23}^\nu c_{12}^\nu}
 e^{-i(\delta_{12}^\nu-\delta_{12}^e)}    \right ] \ .
   \label{phi}
\end{eqnarray}
Therefore, the leptonic CP violating phase $\delta_{\rm PMNS}$ is given as
\begin{eqnarray}
\delta_{\rm PMNS}=\delta_{13}-\delta_{23}-\delta_{12}=
\phi -2\pi\equiv \phi \ .
\label{deltaPMNS}
\end{eqnarray}
Thus, the leptonic CP violating phase is given by $\delta_{12}^e-\delta_{12}^\nu$,
which is similar to $\delta_{CKM}$ of the quark sector in Eq.(\ref{delta12}).
It is remarked that $\delta_{\rm PMNS}$ is given by the  replacement of 
  $d \rightarrow \nu$  and  $u \rightarrow e$ in $\delta_{\rm CKM}$. 
  This situation is understandable since
  the CKM matrix is $V_{uL}^\dagger V_{dL}$ while the PMNS matrix is $V_{eL}^\dagger V_{\nu L}$. 
 
\section{Linking leptons  to quarks}
 \subsection{GUT relation between  quarks and leptons}

 The GUT models 
 relate $\theta_{12}^e$ to the quark mixing angle. 
Especially,  $\theta_{12}^e\simeq \theta_{C}$
with $\theta_{C}$ being the Cabibbo angle,
 comes from the GUT relations between the down-type Yukawa matrix $Y_d$ and the charged lepton
Yukawa matrix $Y_e$, and then one obtains the successful relation 
 $\theta_{13}\simeq \theta_C/\sqrt{2}$ from  Eq.(\ref{prediction13})
\cite{Antusch:2012fb,Antusch:2013kna,Antusch:2011qg,Marzocca:2011dh}.
 
 In order to link the lepton mass matrices to the quark mass matrices,
 we consider the SO(10) GUT, which is broken down to the SM  gauge group 
through the Pati-Salam symmetry \cite{Bando:2003wb,Bando:2004hi}. 
In this setup,
the  quark Yukawa matrices $Y_d$, $Y_u$, the charged lepton  Yukawa matrix $Y_e$
and the Dirac neutrino Yukawa matrix $Y_\nu$ are symmetric ones, and 
 $Y_d=Y_e$ and $Y_u=Y_\nu$ are guaranteed
except for the group theoretical Clebsch-Gordan (CG) factors in each element, which depends on the GUT operators.
 The left-handed Majorana neutrino mass matrix is not directly obtained 
from the up-quark Yukawa matrix $Y_u$
 since it is related with the Dirac neutrino mass matrix with the seesaw mechanism, 
and   we therefore have some freedom
coming from the right-handed neutrino mass matrix $M_R$. 
The remarkable difference of the mixing angles between the CKM matrix and the PMNS matrix
is due to the flavor structure in the right-handed neutrino mass matrix $M_R$. 

Now, one   can take \cite{Antusch:2011qg,Marzocca:2011dh}
\begin{equation}
 \theta_{12}^e= \theta_{12}^d \ ,
 \label{mixinged}
 \end{equation}
 which leads to   $\theta_{13}\simeq \theta_C/\sqrt{2}$.
 This relation was examined  by the renormalization group running \cite{Antusch:2012fb},
 and then, it is justified to be  a good approximation at low energies.
 Therefore, we use it as the low energy relation in our work.
 
As well as the mixing angles,
 the  phases of the Yukawa matrix elements are also related with each other
 in both quark and lepton sectors.
Since  $Y_d=Y_e$ and $Y_u=Y_\nu$ except for CG factors, 
we can discuss  the  relation of the Dirac phases
 between the quark and the lepton sectors.
 In order to simplify the discussion, we take the real base for 
  $Y_u$ and $Y_\nu$, that is  $\delta_{12}^u=\delta_{12}^\nu=0$
 and $\delta_{23}^u=\delta_{23}^\nu=0$.
 One can take this base in general as follows. 
 Through multiplying both side of  Eq.(\ref{qmix23}), 
 Eq.(\ref{qmix13}) and    Eq.(\ref{qmix12}) by the phase factors  $e^{i\delta_{23}^u}$,
$e^{i(\delta_{12}^u+\delta_{23}^u)}$ and $e^{i\delta_{12}^u}$, respectively,
 the  phases $e^{i\delta_{12}^u}$ and $e^{i\delta_{23}^u}$ are removed 
in the up-quark mixing, while the down-quark phases turn to
$\delta_{ij}^d-\delta_{ij}^u$ as seen in the right-hand sides.
The phases of the left-hand sides can be again redefined as $\delta_{ij}$.
In the lepton sector, the same discussion is available 
 in  Eqs.(\ref{lmix23}),  (\ref{lmix13}) and    (\ref{lmix12}).
 
 Due to $Y_d=Y_e$ with the real $Y_u$ and $Y_\nu$, we obtain 
\begin{equation}
\delta_{12}^e   = \delta_{12}^d \ .
\label{phaserelation}
\end{equation}
The neutrino flavor mixing is much different from the up-quark flavor mixing
because of the seesaw mechanism. If the right-handed neutrino mass matrix $M_R$
is real apart from the Majorana phases in this base, Dirac phases do not appear
in the neutrino mixing matrix. The CP violating Dirac phases appear
only in the charged lepton sector.
A typical example is  the tri-bimaximal mixing, which
is the  real mixing with $\theta_{13}^\nu=0$ \cite{Harrison:2002er,Harrison:2002kp}.
Then, the leptonic  CP violating phase comes from the charged lepton sector
as well as  the non-vanishing $\theta_{13}^\nu$.

Taking account of the phase redefinition, 
we rewrite Eq.(\ref{phaserelation})  as 
\begin{equation}
\delta_{12}^e - \delta_{12}^\nu  = \delta_{12}^d- \delta_{12}^u \ ,
\label{linking-phase}
\end{equation}
which is the link of the CP violating phase between quarks and leptons.
By using this relation, we  predict the CP violating phase of leptons,  $\delta_{\rm PMNS}$.

\subsection{Predicting CP violating in  lepton sectors}
 Let us consider the case, in which the mixing angles and the phases  of the quark and lepton Yukawa matrices  
 satisfy the relations in  Eqs. (\ref{mixinged}) and (\ref{linking-phase}), respectively.
Then, the $\delta_{\rm PMNS}$ is given in terms of $\phi_2(\alpha)$  
by using Eq.(\ref{deltaPMNS}):
\begin{eqnarray}
\delta_{\rm PMNS}=
 \phi
= {\rm Arg} \left[1- \frac{s_{12}^\nu}{\theta_{12}^e c_{23}^\nu c_{12}^\nu}
 e^{i\phi_2(\alpha)}    \right ] \ .
\label{CPprediction}
\end{eqnarray}
We can calculate $\delta_{\rm PMNS}$  without assuming 
the flavor structure of the neutrino sector
since the values of  $\theta_{23}^\nu$ and  $\theta_{12}^\nu$  are  obtained
 by the experimental data as seen  in Eqs. (\ref{prediction23}) and  (\ref{prediction12}),
while  $\theta_{12}^e$ is fixed by Eq.(\ref{mixinged}).
For input data in our calculations, we use the results of the global analysis of the neutrino oscillation experiments for $\theta_{12}$, $\theta_{23}$ and  $\theta_{13}$  
\cite{Gonzalez-Garcia:2014bfa,Tortola:2012te,Fogli:2012ua}.

\begin{figure}[b!]
\begin{minipage}[]{0.45\linewidth}
\vspace{4 mm}
\includegraphics[width=7.5cm]{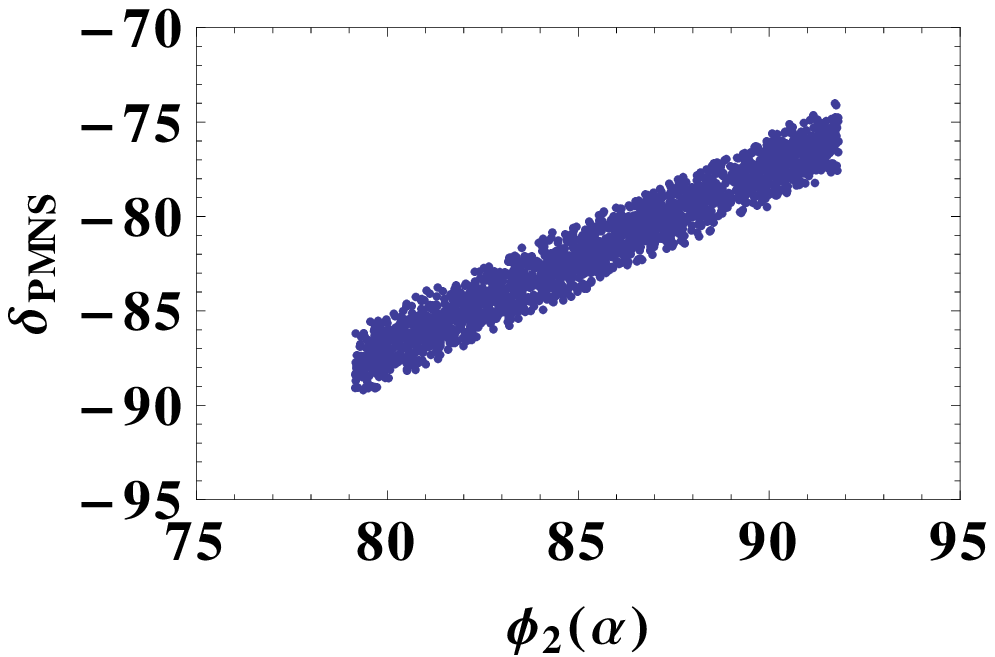}
\caption{Predicted $\delta_{PMNS}$ vs.$\phi_2(\alpha)$.}
\label{fig5}
\end{minipage}
\hspace{5mm}
\begin{minipage}[]{0.45\linewidth}
\vskip - 3 mm
\includegraphics[width=7.5cm]{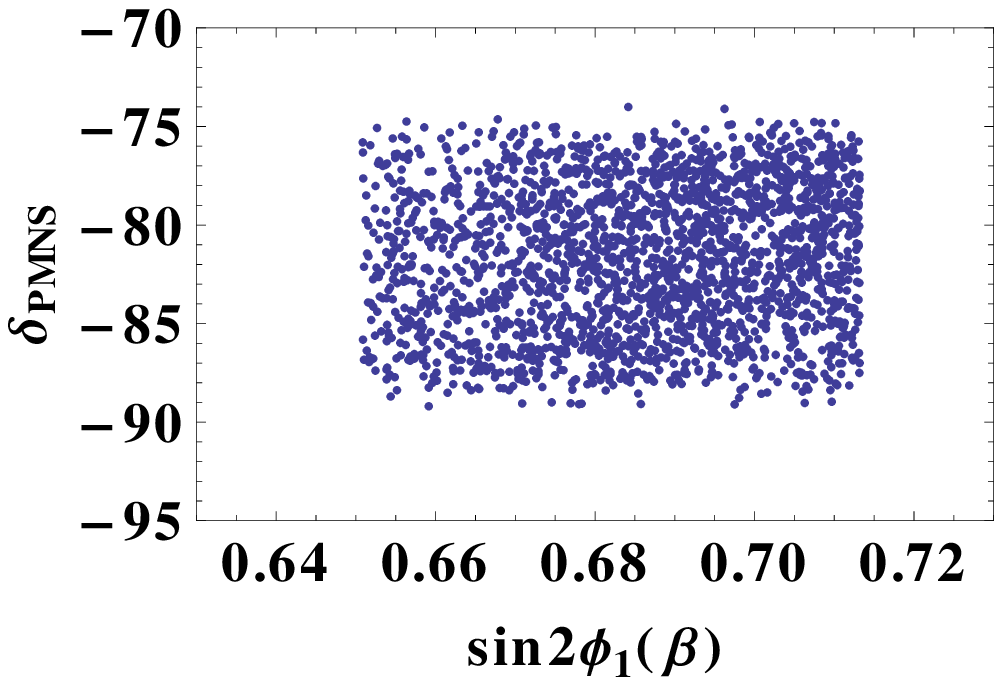}
\caption{Predicted $\delta_{PMNS}$ vs. $\sin 2\phi_1(\beta)$.}
\label{fig6}
\end{minipage}
\end{figure}
\begin{figure}[h!]
\begin{minipage}[]{0.45\linewidth}
\vspace{4 mm}
\includegraphics[width=7.5cm]{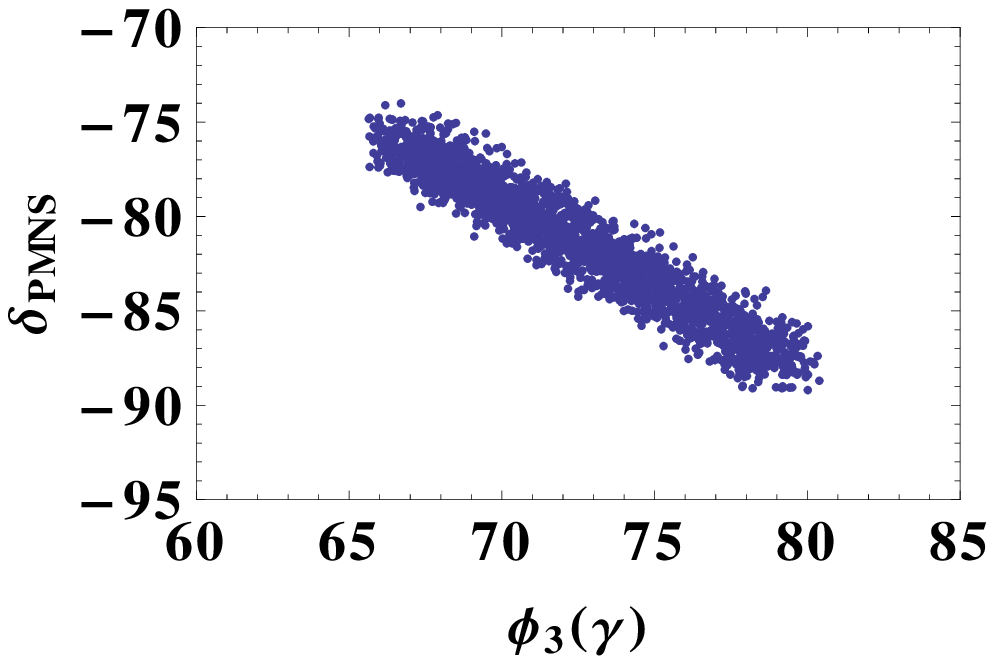}
\caption{Predicted $\delta_{PMNS}$ vs. $\phi_3(\gamma)$.}
\label{fig7}
\end{minipage}
\hspace{5mm}
\begin{minipage}[]{0.45\linewidth}
\vskip - 3 mm
\includegraphics[width=7.5cm]{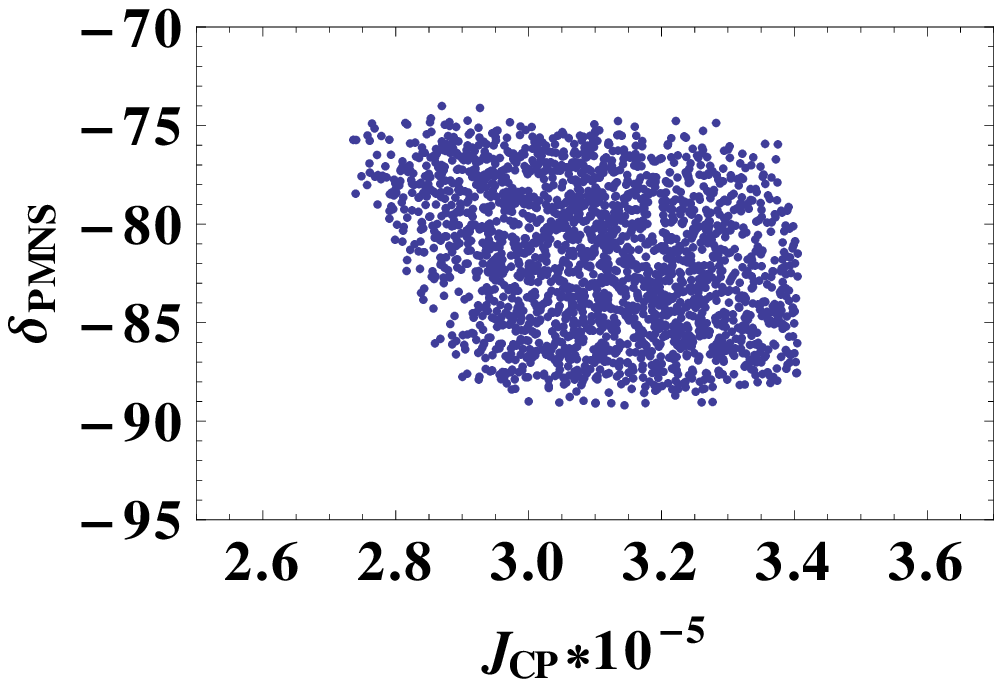}
\caption{Predicted $\delta_{PMNS}$ vs. $J_{CP}$.}
\label{fig8}
\end{minipage}
\end{figure}

Now we predict $\delta_{\rm PMNS}$ 
in terms of the CP violating phase of the quarks  by using 
 Eq.(\ref{CPprediction}).
We obtain   $\delta_{\rm PMNS}$ in the region of  $-74^\circ\sim -89^\circ$
 which is  consistent with the recent  T2K results with $90\%$ C.L. \cite{Abe:2013hdq}; 
$-1.18\pi <\delta _{CP}<0.15\pi ~(-0.91\pi <\delta _{CP}<-0.08\pi )$ 
for the normal (inverted) hierarchy of neutrinos.
The predict  $\delta_{\rm PMNS}$ is  the opposite sign against   $\delta_{\rm CKM}$, which 
 is due to that $\delta_{\rm PMNS}$ is given by  replacement of 
  $d \rightarrow \nu$  and  $u \rightarrow e$ in $\delta_{\rm CKM}$. 

We show the predicted $\delta_{\rm PMNS}$ versus $\phi_2(\alpha)$, $\sin 2\phi_1(\beta)$,
$\phi_3(\gamma)$ and $J_{CP}$ in Figures 5, 6, 7 and 8, respectively.
As seen in Figure 5,  $\delta_{\rm PMNS}$ has a linear dependence of $\phi_2(\alpha)$ 
with somewhat uncertainty, which is due to the error
 of input data, $\theta_{12}$ and $\theta_{23}$, 
 as seen in  Eqs. (\ref{prediction23}), (\ref{prediction12}) and (\ref{CPprediction}).
 The more precise data of $\phi_2(\alpha)$ in the $B^0$ meson system
 gives the more accurate prediction of  $\delta_{\rm PMNS}$.
 If the $\phi_2(\alpha)$ is just a right angle,  $\delta_{\rm PMNS}$
should be $-75^\circ\sim -80^\circ$.

On the other hand, $\delta_{\rm PMNS}$ is insensitive to $\sin 2\phi_1(\beta)$.
As seen in Eq.(\ref{triangle}), $\sin 2\phi_1(\beta)$ depends  on  $\theta_{12}^u/\theta_{12}^d$, where
 the uncertainty of  $\theta_{12}^d$ and   $\theta_{12}^u$  comes from  the experimental
 error-bar of the CKM matrix elements $|U_{ub}|$, $|U_{cb}|$ 
and $\phi_3(\gamma)$ as seen in  Eq.(\ref{d12}). 
 Therefore,  the  $\sin 2\phi_1(\beta)$ dependence of $\delta_{\rm PMNS}$ 
 will be seen  after the  more precise determination of the CKM matrix elements $|U_{ub}|$, 
 $|U_{cb}|$ and $\phi_3(\gamma)$ in the future experiments.

 It is  found that $\delta_{\rm PMNS}$ is sensitive to $\phi_3(\gamma)$, which 
 also depends on  $\theta_{12}^d/\theta_{12}^u$. 
 In this case,  since $\theta_{12}^d$ is given for the fixed  $\phi_3(\gamma)$,
  the uncertainty of  $\theta_{12}^d$ is significantly reduced.
The predicted $\delta_{\rm PMNS}$ is   insensitive to $J_{CP}$.


\vskip -1 cm
\section{Summary}

We have discussed the linking between the CP violating phase of the  lepton sector
and  the unitarity triangle of the quark sector.
The   1-3 texture zeros (negligible 1-3 mixing) 
in the up- and the down-quark  mass matrices have given us an interesting relation 
between the Yukawa phase and  CP violating measure $\phi_2(\alpha)$ \cite{Antusch:2009hq}.
We have examined  this relation by using the recent data of 
three  angles of the unitarity triangle in the $B^0$ meson system and
 the CP violating measure $J_{CP}$.
We have found that the CP violation is successfully  expressed in terms of
$\delta_{12}^d-\delta_{12}^u$ and $\theta_{12}^d/\theta_{12}^u$,
which come from the quark mass matrices. 
 
We have extended this approach by considering 
 the $SO(10)$ GUT including the Pati-Salam symmetry, which
links the Yukawa couplings of the quark sector  to the one of the lepton sector. 
Then, there is a simple  relation of the CP violating Dirac phases 
between both quark and lepton sectors as well as the mixing angles.
We have predicted  the leptonic CP violating phase 
in terms of the angle of the unitaity triangle of  quarks.
 The leptonic CP violating phase $\delta_{\rm PMNS}$ is in the region $-74^\circ\sim -89^\circ$,
 which is the opposite sign against   $\delta_{\rm CKM}$.
 The predicted value is the consistent with the recent  T2K results.
Our predicted $\delta_{\rm PMNS}$ is sensitive to  $\phi_2(\alpha)$ and 
 $\phi_3(\gamma)$, on the other hand, insensitive to  $\sin 2\phi_1(\beta)$ and $J_{CP}$.
Our predictions will be  tested in the future neutrino experiments
and the Belle-II experiment.

\vspace{1 cm}
\noindent
{\bf Acknowledgment}
\vspace{0.5 cm}

M.T.and K.Y  are  supported by JSPS
 Grand-in-Aid for Scientific Research,  No.24654062. and No.25-5222, respectively.
 M.T would like to thank Sin Kyu Kang for helpful discussion.

\newpage
\appendix{}
\section*{Appendix : General mixing  of Quarks and Leptons}

The CKM matrix is given in general as:
\begin{eqnarray}
U'_{\rm CKM} =U_{12}^{uL\dagger} U_{13}^{uL\dagger} U_{23}^{uL\dagger} 
   U_{23}^{dL} U_{13}^{dL} U_{12}^{dL} \ .
\end{eqnarray}
Then, we have mixing sum rules:
\begin{eqnarray}
&&\theta_{23} e^{-i\delta_{23}}=(s_{23}^d e^{-i\delta_{23}^d}
  -s_{23}^u e^{-i\delta_{23}^u})+ s_{12}^u e^{i\delta_{12}^u} 
(s_{13}^d e^{-i\delta_{13}^d}-s_{13}^u e^{-i\delta_{13}^u}) \ , 
 \nonumber \\
&&\theta_{13} e^{-i\delta_{13}}=-s_{12}^u  e^{-i\delta_{12}^u}
(s_{23}^d e^{-i\delta_{23}^d} -s_{23}^u e^{-i\delta_{23}^u}) 
+(s_{13}^d e^{-i\delta_{13}^d}-s_{13}^u e^{-i\delta_{13}^u}) \ , 
   \label{quarksumA} \\ 
&&\theta_{12} e^{-i\delta_{12}}=(s_{12}^d e^{-i\delta_{12}^d}
  -s_{12}^u e^{-i\delta_{12}^u})+s_{13}^u e^{-i\delta_{13}^u} 
(s_{23}^d e^{-i\delta_{23}^d}
  -s_{23}^u e^{-i\delta_{23}^u}) \ , \nonumber
\end{eqnarray}
where we keep the second  order of  the small mixing angles
 by putting  $c_{ij}^{u,d}\simeq 1$ and  $s_{ij}^{u,d}\ll 1$.
 
 The angle of the CP violation $\phi_2(\alpha)$ is  defined in terms of the CKM matrix elements as
\begin{eqnarray}
\phi_2 (\alpha) = {\text Arg} \left [-\frac{U_{td}U_{tb}^*}{U_{ud}U_{ub}^*} \right ]
\simeq   {\text Arg} \left [-\frac{U_{td}}{U_{ub}^*} \right ] \ , 
\end{eqnarray}
where  $U_{ub}$ is given by $\theta_{13} e^{-i\delta_{13}}$
in Eq.(\ref{quarksumA}).
On the other hand, $U_{td}$ is expressed as 
\begin{eqnarray}
U_{td}=s_{12}^d  e^{i\delta_{12}^d}
(s_{23}^d e^{i\delta_{23}^d}-s_{23}^u e^{i\delta_{23}^u}) 
+(s_{13}^u e^{i\delta_{13}^u}-s_{13}^d e^{i\delta_{13}^d}) \ . 
\end{eqnarray}
Then,  $\phi_2(\alpha)$ is obtained as follows:
\begin{eqnarray}
\phi_2 (\alpha) = {\text Arg} \left [
\frac{s_{12}^d  e^{i\delta_{12}^d}
(s_{23}^d e^{i\delta_{23}^d}-s_{23}^u e^{i\delta_{23}^u}) 
+(s_{13}^u e^{i\delta_{13}^u}-s_{13}^d e^{i\delta_{13}^d})}
{s_{12}^u  e^{i\delta_{12}^u}
(s_{23}^d e^{i\delta_{23}^d} -s_{23}^u e^{i\delta_{23}^u}) 
+(s_{13}^u e^{i\delta_{13}^u}-s_{13}^d e^{i\delta_{13}^d})}
 \right ] \ . 
 \label{alphaA}
\end{eqnarray}
We obtain easily  $\phi_2 (\alpha)=\delta_{12}^d-\delta_{12}^u$
if $s_{13}^u=s_{13}^d=0$ is put.
In the case of non-vanishing  $s_{13}^u$ and $s_{13}^d$, $\phi_2 (\alpha)$ depends on
the extra parameters of the (1-3) mixing in additon to
 $\delta_{12}^d-\delta_{12}^u$.
Even if  $s_{13}^d$  has a non-vanishing value with $s_{13}^u=0$,
this situation is not changed, and vice versa.
Thus, we lose the predictive power for the CP violating phase unless $s_{13}^u=s_{13}^d=0$.

Instead  of  $s_{13}^u=s_{13}^d=0$, 
if we put the  $s_{12}^u=s_{12}^d=0$ or  $s_{23}^u=s_{23}^d=0$ 
 we get $\phi_2 (\alpha)=0$  as seen in  Eq.(\ref{alphaA}),
which is inconsistent with the experimental data of the CKM mixing.
In conclusion, $s_{13}^u=s_{13}^d=0$ is the minimal choice
 in the standpoint of the texture zeros to discuss
the CP violation of the CKM mixing.

Let us discuss  the PMNS matrix, which is given as,
\begin{eqnarray}
U'_{\rm PMNS} =U_{12}^{e L\dagger} U_{13}^{e L\dagger} U_{23}^{e L\dagger} 
U_{23}^{\nu L} U_{13}^{\nu L} U_{12}^{\nu L} \ .
\end{eqnarray}
Taking $s_{ij}^{e}\ll 1$ in the charged leptons and  the mixing angles with order one in the neutrinos,
we obtain
\begin{eqnarray}
&&c_{13}s_{23} e^{-i\delta_{23}}=c_{13}^\nu s_{23}^\nu e^{-i\delta_{23}^\nu} 
-s_{23}^e c_{23}^\nu c_{13}^\nu e^{-i\delta_{23}^e}
+s_{12}^e s_{13}^\nu  e^{-i(\delta_{13}^e-\delta_{12}^e)}   \ ,  \nonumber
 \\
&&s_{13} e^{-i\delta_{13}}=-s_{12}^e c_{13}^\nu e^{-i\delta_{12}^e}
(s_{23}^\nu e^{-i\delta_{23}^\nu} -s_{23}^e c_{23}^\nu e^{-i\delta_{23}^e}) 
+s_{13}^\nu e^{-i\delta_{13}^\nu} -s_{13}^e c_{13}^\nu c_{23}^\nu  e^{-i\delta_{13}^e} \ , 
\label{lsumA} \\
&&c_{13}s_{12} e^{-i\delta_{12}}= c_{13}^\nu s_{12}^\nu e^{-i\delta_{12}^\nu}
  -s_{12}^e c_{12}^\nu c_{23}^\nu    e^{-i\delta_{12}^e}  +
  s_{13}^e  s_{23}^\nu c_{12}^\nu  e^{-i(\delta_{13}^e-\delta_{23}^\nu)}  \ . \nonumber
\end{eqnarray}
Since the observed $s_{23}$ and  $s_{12}$  are large, 
we cannot take $s_{23}^\nu=0$ and $s_{12}^\nu=0$
if the charged lepton mixing angles is assumed to be  order of the quark ones
through the GUT model.  Therefore, the input of $s_{13}^\nu=0$
in addition to  $s_{13}^e=0$ is a reasonable one.
If we take  $s_{13}^\nu\not =0$, it is  an extra parameter as seen in Eq.(\ref{lsumA}). 
Then, we lose the predictive power for $\theta_{13}$ and the CP violating phase $\delta_{\rm PMNS}$.



\begin{thebibliography}{110}


  
 
  

\bibitem{Abe:2013hdq}
  K.~Abe {\it et al.}  [T2K Collaboration],
  Phys.\ Rev.\ Lett.\  {\bf 112} (2014) 061802
  [arXiv:1311.4750 [hep-ex]].
 

\bibitem{An:2012eh}
  F.~P.~An {\it et al.}  [DAYA-BAY Collaboration],
  Phys.\ Rev.\ Lett.\  {\bf 108} (2012) 171803
  [arXiv:1203.1669 [hep-ex]].
  
\bibitem{Ahn:2012nd}
  J.~K.~Ahn {\it et al.}  [RENO Collaboration],
 Phys.\ Rev.\ Lett.\  {\bf 108} (2012) 191802  [arXiv:1204.0626 [hep-ex]].
 
\bibitem{Adamson:2013whj}
  P.~Adamson {\it et al.}  [MINOS Collaboration],
 Phys.\ Rev.\ Lett.\  {\bf 110} (2013) 25,  251801  [arXiv:1304.6335 [hep-ex]].
 
\bibitem{Abe:2014lus}
  Y.~Abe {\it et al.}  [Double Chooz Collaboration],
 arXiv:1401.5981 [hep-ex].
\bibitem{Kang:2000sb}
  S.~K.~Kang, C.~S.~Kim and J.~D.~Kim,
  Phys.\ Rev.\ D {\bf 62} (2000) 073011
  [hep-ph/0004020].
  
\bibitem{Fukugita:2001rk}
  M.~Fukugita and M.~Tanimoto,
  Phys.\ Lett.\ B {\bf 515} (2001) 30
  [hep-ph/0107082].
  
\bibitem{Giunti:2002pp}
  C.~Giunti and M.~Tanimoto,
  Phys.\ Rev.\ D {\bf 66} (2002) 113006
  [hep-ph/0209169].

\bibitem{Xing:2002sw}
  Z.~-z.~Xing,
  Phys.\ Lett.\ B {\bf 533} (2002) 85
  [hep-ph/0204049].

\bibitem{Adhikary:2006jx}
  B.~Adhikary and A.~Ghosal,
  Phys.\ Rev.\  D {\bf 75}, 073020 (2007)
  [arXiv:hep-ph/0609193].

\bibitem{Adhikary:2008au}
  B.~Adhikary and A.~Ghosal,
  Phys.\ Rev.\  D {\bf 78}, 073007 (2008)
  [arXiv:0803.3582 [hep-ph]].

\bibitem{Branco:2011zb}
  G.~C.~Branco, R.~G.~Felipe and F.~R.~Joaquim,
  Rev.\ Mod.\ Phys.\  {\bf 84} (2012) 515
  [arXiv:1111.5332 [hep-ph]].
  
\bibitem{Branco:2012vs} 
  G.~C.~Branco, R.~Gonzalez Felipe, F.~R.~Joaquim and H.~Serodio,
  Phys.\ Rev.\ D {\bf 86}, 076008 (2012)
  [arXiv:1203.2646 [hep-ph]].

\bibitem{Ahn:2012tv}
  Y.~H.~Ahn and S.~K.~Kang,
  Phys.\ Rev.\ D {\bf 86} (2012) 093003
  [arXiv:1203.4185 [hep-ph]].

 
 \bibitem{Ishimori:2012sw}
  H.~Ishimori, S.~Khalil and E.~Ma,
  Phys.\ Rev.\ D {\bf 86} (2012) 013008
  [arXiv:1204.2705 [hep-ph]].
  
\bibitem{Ishimori:2012fg}
  H.~Ishimori and E.~Ma,
  Phys.\ Rev.\ D {\bf 86} (2012) 045030
  [arXiv:1205.0075 [hep-ph]].


 
\bibitem{Rodejohann:2012cf}
  W.~Rodejohann and H.~Zhang,
  Phys.\ Rev.\ D {\bf 86} (2012) 093008
  [arXiv:1207.1225 [hep-ph]].
 
  \bibitem{Marzocca:2013cr}
  D.~Marzocca, S.~T.~Petcov, A.~Romanino and M.~C.~Sevilla,
  JHEP {\bf 1305} (2013) 073
  [arXiv:1302.0423 [hep-ph]].

\bibitem{Ballett:2013wya}
  P.~Ballett, S.~F.~King, C.~Luhn, S.~Pascoli and M.~A.~Schmidt,
  Phys.\ Rev.\ D {\bf 89} (2014) 1,  016016
  [arXiv:1308.4314 [hep-ph]].
  
\bibitem{Ballett:2014dua}
  P.~Ballett, S.~F.~King, C.~Luhn, S.~Pascoli and M.~A.~Schmidt,
  JHEP {\bf 1412} (2014) 122
  [arXiv:1410.7573 [hep-ph]].

\bibitem{King:2014nza}
  S.~F.~King, A.~Merle, S.~Morisi, Y.~Shimizu and M.~Tanimoto,
 arXiv:1402.4271 [hep-ph].


\bibitem{Xing:2014zka}
  Z.~-z.~Xing and S.~Zhou,
  arXiv:1404.7021 [hep-ph].
  
\bibitem{Branco:2014zza}
  G.~C.~Branco, M.~N.~Rebelo, J.~I.~Silva-Marcos and D.~Wegman,
  Phys.\ Rev.\ D {\bf 91} (2015) 013001
  [arXiv:1405.5120 [hep-ph]].

  

\bibitem{Petcov:2014laa}
  S.~T.~Petcov,
  Nucl.\ Phys.\ B {\bf 892} (2015) 400
  [arXiv:1405.6006 [hep-ph]].

  
\bibitem{Kang:2014mka}
  S.~K.~Kang and C.~S.~Kim,
  Phys.\ Rev.\ D {\bf 90} (2014) 7,  077301
  [arXiv:1406.5014 [hep-ph]].
  
\bibitem{Shimizu:2014ria}
  Y.~Shimizu, M.~Tanimoto and K. Yamamoto,
 Mod. Phys. Lett. A, {\bf 30} (2015) 1550002  [arXiv:1405.1521 [hep-ph]].
  

 
\bibitem{Girardi:2014faa}
  I.~Girardi, S.~T.~Petcov and A.~V.~Titov,
  arXiv:1410.8056 [hep-ph].
   
\bibitem{Kang:2015xfa}
  S.~K.~Kang and M.~Tanimoto,
  arXiv:1501.07428 [hep-ph].
  
\bibitem{Antusch:2009hq}
  S.~Antusch, S.~F.~King, M.~Malinsky and M.~Spinrath,
  Phys.\ Rev.\ D {\bf 81} (2010) 033008
  [arXiv:0910.5127 [hep-ph]].
 
 

\bibitem{Antusch:2012fb}
  S.~Antusch, C.~Gross, V.~Maurer and C.~Sluka,
  Nucl.\ Phys.\ B {\bf 866} (2013) 255
  [arXiv:1205.1051 [hep-ph]].
  
\bibitem{Antusch:2013kna}
  S.~Antusch, C.~Gross, V.~Maurer and C.~Sluka,
  Nucl.\ Phys.\ B {\bf 877} (2013) 772
  [arXiv:1305.6612 [hep-ph]].
  
\bibitem{Antusch:2011qg}
  S.~Antusch and V.~Maurer,
 Phys.\ Rev.\ D {\bf 84} (2011) 117301  [arXiv:1107.3728 [hep-ph]].
 
\bibitem{Marzocca:2011dh}
  D.~Marzocca, S.~T.~Petcov, A.~Romanino and M.~Spinrath,
  JHEP {\bf 1111} (2011) 009
  [arXiv:1108.0614 [hep-ph]].
  
\bibitem{BhupalDev:2011gi}
  P.~S.~Bhupal Dev, R.~N.~Mohapatra and M.~Severson,
  Phys.\ Rev.\ D {\bf 84} (2011) 053005
  [arXiv:1107.2378 [hep-ph]].
  
\bibitem{BhupalDev:2012nm}
  P.~S.~Bhupal Dev, B.~Dutta, R.~N.~Mohapatra and M.~Severson,
  Phys.\ Rev.\ D {\bf 86} (2012) 035002
  [arXiv:1202.4012 [hep-ph]].

  


\bibitem{Bando:2003wb}
  M.~Bando, S.~Kaneko, M.~Obara and M.~Tanimoto,
  Phys.\ Lett.\ B {\bf 580} (2004) 229
  [hep-ph/0309310].
  
\bibitem{Bando:2004hi}
  M.~Bando, S.~Kaneko, M.~Obara and M.~Tanimoto,
  Prog.\ Theor.\ Phys.\  {\bf 112} (2004) 533
  [hep-ph/0405071].
  




\bibitem{Beringer:1900zz}
  J.~Beringer {\it et al.}  [Particle Data Group Collaboration],
  Phys.\ Rev.\ D {\bf 86} (2012) 010001.

\bibitem{Jarlskog:1985ht}
  C.~Jarlskog,
  Phys.\ Rev.\ Lett.\  {\bf 55} (1985) 1039.

\bibitem{Maki:1962mu}
  Z.~Maki, M.~Nakagawa and S.~Sakata,
 Prog.\ Theor.\ Phys.\  {\bf 28} (1962) 870.

\bibitem{Pontecorvo:1967fh}
  B.~Pontecorvo,
  Sov.\ Phys.\ JETP {\bf 26} (1968) 984
   [Zh.\ Eksp.\ Teor.\ Fiz.\  {\bf 53} (1967) 1717].
  
  
  
\bibitem{Harrison:2002er}
  P.~F.~Harrison, D.~H.~Perkins, W.~G.~Scott,
  Phys.\ Lett.\ B {\bf 530 } (2002)  167
  [hep-ph/0202074].


\bibitem{Harrison:2002kp}
  P.~F.~Harrison, W.~G.~Scott,
  Phys.\ Lett.\ B {\bf 535 } (2002)  163-169
  [hep-ph/0203209].
 
\bibitem{Gonzalez-Garcia:2014bfa}
  M.~C.~Gonzalez-Garcia, M.~Maltoni and T.~Schwetz,
  JHEP {\bf 1411} (2014) 052
  [arXiv:1409.5439 [hep-ph]].
  

\bibitem{Tortola:2012te}
  D.~V.~Forero, M.~Tortola and J.~W.~F.~Valle,
  arXiv:1205.4018 [hep-ph].

\bibitem{Fogli:2012ua}
  G.~L.~Fogli, E.~Lisi, A.~Marrone, D.~Montanino, A.~Palazzo and A.~M.~Rotunno,
  Phys.\ Rev.\ D {\bf 86} (2012) 013012
  [arXiv:1205.5254 [hep-ph]].


  

  



\end{thebibliography}
\end{document}